\documentclass[twocolumn,showpacs,preprintnumbers,amsmath,amssymb]{revtex4}

\usepackage{graphicx}
\usepackage{dcolumn}
\usepackage{bm}

\begin{document}

\preprint{}

\title{Revealing the Exciton Fine Structure in PbSe Nanocrystal Quantum Dots}

\author{R. D. Schaller$^{1}$, S. A. Crooker$^{2}$, D. A. Bussian$^{1}$, J. M. Pietryga$^{1}$, J. Joo$^{1}$, V. I. Klimov$^{1}$}

\affiliation{$^1$Chemistry Division, Los Alamos National Laboratory, Los Alamos, NM 87545}
\affiliation{$^2$National High Magnetic Field Laboratory, Los Alamos, NM 87545}

\date{\today}
\begin{abstract}
We measure the photoluminescence (PL) lifetime, $\tau$, of excitons
in colloidal PbSe nanocrystals (NCs) at low temperatures to 270~mK
and in high magnetic fields to 15~T. For all NCs (1.3-2.3~nm radii),
$\tau$ increases sharply below 10~K but saturates by 500~mK. In
contrast to the usual picture of well-separated ``bright" and
``dark" exciton states (found, e.g., in CdSe NCs), these dynamics
fit remarkably well to a system having two exciton states with
comparable - but small - oscillator strengths that are separated by
only 300-900 $\mu$eV. Importantly, magnetic fields reduce $\tau$
below 10~K, consistent with field-induced mixing between the two
states. Magnetic circular dichroism studies reveal exciton
\emph{g}-factors from 2-5, and magneto-PL shows $>$10\% circularly
polarized emission.

\end{abstract}
\maketitle
Lead-salt semiconductor nanocrystals (NCs) based on PbS, PbSe, and
PbTe are actively studied due to their potential impact on important
technologies including telecommunication-wavelength optoelectronics
(they exhibit size-tunable emission spanning the near- to
mid-infrared \cite{Wise, Murray, Pietryga}) and enhanced solar
energy conversion (multiple excitons can be formed upon absorption
of single photons \cite{Schaller,Ellingson}). Other distinguishing
characteristics include high photoluminescence (PL) quantum yields
\cite{Wehrenberg}, very long exciton radiative lifetimes ($\tau$)
\cite{Wehrenberg, Du, Kigel, Oron} and the ability to form
core-shell heterostructures \cite{Lifshitz}. However, in contrast to
their counterparts that emit at visible wavelengths (e.g., CdSe or
CdS NCs), the fundamental electronic structure of lead-salt NCs
remains an open question, despite more than a decade of experiments
and often-conflicting theoretical consideration \cite{Kang, Allan,
An1, An2}.

Electron-hole exchange interactions, crystal structure, band
symmetry, spin-orbit coupling, and shape anisotropy all influence
the underlying level ordering and oscillator strengths of band-edge
excitons in colloidal NCs \cite{Nirmal, Efros, Chamarro, LeThomas}.
Understanding this ``exciton fine structure" is especially important
in NCs because it governs both absorptive and emissive optical
properties. For example, in the common case of CdSe NCs these
properties combine to split the \emph{1S} band-edge exciton into the
now well-established fine structure of five distinct levels, wherein
the lowest energy state is an optically-forbidden ``dark" exciton
that lies as much as $\Delta$=2-15~meV below the nearest
optically-allowed ``bright" exciton. Crucially, proof of dark
excitons in CdSe NCs relied on measurements of PL and PL decays at
low temperatures ($T$$<$2K, so that $k_B T < \Delta$) and in high
magnetic fields ($B$$>$10~T, so that the magnetic energy $g_{ex}
\mu_B B \simeq \Delta$)\cite{Nirmal, Crooker, Furis}.  These studies
revealed surprisingly long PL lifetimes at low temperature ($\sim$1
$\mu$s) that shortened with applied field, consistent with
optically-dark excitons that gain oscillator strength due to
field-induced mixing with bright states.

In contrast with wide-gap semiconductors like CdSe (which have
wurtzite/zincblende structure and direct bandgaps at the
Brillouin-zone center $\Gamma$-point), lead-salt semiconductors such
as PbSe are narrow-gap materials having rock-salt crystal structure
and direct gaps at the four-fold degenerate $L$-point at the
Brillouin zone edge. In further distinction, electrons and holes in
PbSe exhibit very similar masses and giant $g$-factors
($|g|$$\sim$30) due to strong spin-orbit coupling \cite{Bangert}. As
such, exciton fine structure in lead-salt NCs is expected to be
rather different, and several theoretical studies have been reported
for PbSe NCs:  Four-band envelope wavefunction methods first
suggested 1-5 meV exchange energies and predicted an
optically-allowed exciton ground state \cite{Kang}, whereas
tight-binding calculations anticipated that the nominally degenerate
$L$-points are split in NCs by tens of meV by intervalley coupling
\cite{Allan}. More recently, empirical pseudopotential approaches
\cite{An2} suggested that PbSe NCs possess a single
optically-forbidden exciton ground state that lies $\Delta$=2-17~meV
below a 3-fold degenerate manifold of optically-allowed exciton
levels.

Experimentally, the PL decay time ($\tau$) from PbSe NCs was first
reported to be hundreds of nanoseconds at room temperature \cite{Du,
Wehrenberg}, which is long compared to CdSe ($\tau$=20~ns). As a
possible explanation, dark exciton ground states and large
bright-dark splittings ($\Delta$$>$25 meV) were considered, as were
enhanced dielectric screening effects \cite{Wehrenberg}. Subsequent
studies revealed that $\tau$ increased to 1-5 $\mu$s upon cooling from 200$\rightarrow$50~K, prompting suggestions that
$\Delta$ lies in this energy range ($\sim$$k_B$$\cdot$100~K)
\cite{Kigel}. Moreover, $\tau$ was found to increase yet
\emph{again} down to reported base temperatures of 1.4~K
\cite{Kigel,Oron}, which was attributed to freezing-out of quantized
acoustic phonon modes in the NC (which were thought to assist the
recombination of presumed dark excitons). However, 1.4~K was not
sufficiently low to saturate $\tau$, complicating accurate fitting
of the relevant energy scale.  Further, no studies of $\tau$ in
magnetic field, which were essential for confirming bright and
dark exciton states in CdSe NCs, have been reported to date.

Here we measure PL and PL decay times $\tau$ in infrared-emitting
rocksalt PbSe NCs at temperatures over five times lower than
previously reported (to 270 mK), and also in high magnetic fields to
15~T. We find that $\tau$ increases sharply below 10 K, but
saturates below 500~mK. In marked contrast to the `conventional'
exciton fine structure found in CdSe and other wide-gap NCs (i.e.,
clear dark and bright exciton levels with orders-of-magnitude
different oscillator strengths, well-separated in energy by 2-15
meV), the dynamics in PbSe NCs fit remarkably well to a new
exciton structure containing two \emph{weakly}-emitting states with
\emph{comparable} oscillator strength, that are split by a
surprisingly small energy of only 290-870 $\mu$eV
(\emph{r}=2.3-1.3~nm). This energy scale is much smaller than
recently predicted for PbSe NCs \cite{An2}, and is also much less
than any quantized phonon energy in the NC. Importantly, magnetic
fields reduce $\tau$ only below 10 K, consistent with field-induced
mixing between these two states. Further, magnetic circular
dichroism studies establish the magnetic Zeeman energy of the $1S$
absorbing exciton states in PbSe NCs for the first time ($|g_{ex}|$
ranges from 2-5), and magneto-PL from the emitting excitons reveals
$>$10\% circular polarization, also implicating contributions from
Zeeman-split excitons.

\begin{figure}[tbp]
\includegraphics[width=.48\textwidth]{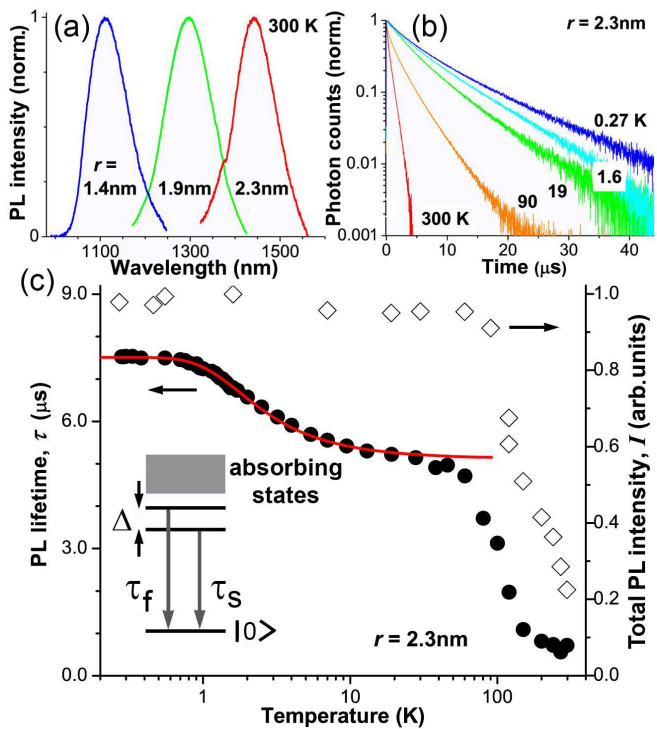}
\caption{(a) PL spectra from PbSe NC ensembles of different size.
(b) PL decays. The lifetime $\tau$ increases as temperature $T$
decreases. (c) Left axis: $\tau$ versus $T$ from 270~mK  to 300~K.
Line is a fit of the low-$T$ data to a two-level model (see inset).
Right axis: The integrated PL emission, $I$. Note \emph{both} $\tau$ and $I$
increase $\sim$fivefold upon cooling to 50~K in these NCs, indicating freeze-out
of competing non-radiative recombination channels. $I$
remains unchanged below 50~K.} \label{fig1}
\end{figure}

Oleic-acid capped PbSe NCs having high quantum yield ($>$20\% at 300K) were
synthesized following Refs. \cite{Murray, Pietryga} and were
dissolved in liquid n-octadecane to form dilute solid-solution films
devoid of inter-NC energy transfer. Fig. 1(a) shows PL spectra from
typical films.  NC radii $r$ are derived from the measured bandgap
\cite{Moreels}. PL decays were measured in the $^3$He insert of a
15~T magnet. The samples were weakly excited by a 635~nm diode laser
delivering 70 ps pulses at 20 kHz (1.5 $\mu$W average power).
Reducing the laser power tenfold at 270~mK did not affect $\tau$,
indicating negligible heating. A 550~$\mu$m diameter optical fiber
delivered the excitation and collected the PL, which was dispersed
in a 0.3~m spectrometer and detected with an InGaAs array or an
InGaAsP photomultiplier tube and photon counting electronics. PL
decays were always recorded at the PL band maximum (which shifts
with temperature \cite{Kigel, Olkhovets}).

Figure 1(b) shows that PL decays from $r$=2.3~nm PbSe NCs are
predominantly single-exponential and become longer with decreasing
temperature. Decay times $\tau$ were extracted from single
exponential fits and are
shown over three orders of magnitude in temperature in Fig. 1(c).  The initial 2 $\mu$s of data are disregarded for $T$$<$100K, to avoid artifacts from exciton cooling or residual inter-dot energy transfer \cite{Furis}.
Also shown (right axis) is the corresponding \emph{total} (time- and
spectrally-integrated) PL intensity, $I$. In qualitative agreement
with prior work \cite{Kigel}, $\tau$=0.7 $\mu$s at 300~K, but
increases to 5 $\mu$s upon cooling to 50~K. Critically, however, we
find that $I$ increases by a similar factor (of about five for this
film) over the same temperature range, below which it remains
constant. A similar correspondence -- namely, that $I$ increases by
approximately the same factor as $\tau$ -- is observed for
\emph{all} NC films upon cooling to 50~K. This result strongly
suggests that non-radiative recombination channels (as opposed to,
e.g., an interplay between dark and bright exciton states) dominates
the PL dynamics above 50~K: As competing non-radiative decay
channels freeze out, both $\tau$ and $I$ should increase
correspondingly. Thus, room temperature PL decay times of order 1
$\mu$s, measured here and in prior works, are likely \emph{not}
representative of the actual radiative exciton lifetime in PbSe NCs.
Rather, these data suggest that $\tau$ measured in the 10-50~K range
provides a more realistic measure.

The most striking aspect of Fig. 1(c) is that $\tau$ remains
relatively constant from 50 to 10~K, but increases markedly yet
again at lower temperatures. No corresponding change of $I$ is
observed, indicating an intrinsic effect consistent with a
redistribution of excitons having near-unity PL quantum yield.
Ultra-low temperatures $<$500~mK are necessary to saturate $\tau$
(at 7.5 $\mu$s in this sample). A key feature of this data is its
remarkable agreement with a simple model of two thermally-populated
emitting exciton levels: a lowest-energy state with slower decay
time $\tau_s$ and a higher-lying state with somewhat faster lifetime
$\tau_f$, separated in energy by $\Delta$ (see inset). In this case,
the temperature-dependent lifetime is readily given by $\tau^{-1}
(T)= (\tau_s^{-1} + \tau_f^{-1} e^{-\Delta/k_B T})/(1 +
e^{-\Delta/k_B T})$. Importantly, the clear saturation of $\tau$
below 500~mK effectively fixes $\tau_s$, permitting $\Delta$ to be
fit with high accuracy. In these 2.3~nm radius NCs, we find that
$\Delta$ is surprisingly small -- only 290$\pm$6 $\mu$eV. Although
previous studies to 1.4~K ascribed the low-$T$ upturn of $\tau$ with
freezeout of quantized $l$=2 acoustic phonons, we note that 290
$\mu$eV is over three times less than the smallest acoustic phonon
energy in PbSe NCs of this size, computed numerically
\cite{Migliori} or using Lamb theory. Thus, this energy scale
$\Delta$ likely reflects an intrinsic splitting of the lowest two
states in the exciton fine structure of PbSe NCs.  Moreover,
$\tau_s$ and $\tau_f$ are similar (7.5 and 3.8 $\mu$s), in strong
contrast to the well-defined dark and bright excitons in CdSe NCs
whose lifetimes typically differ by two orders of magnitude.

\begin{figure}[tbp]
\includegraphics[width=.48\textwidth]{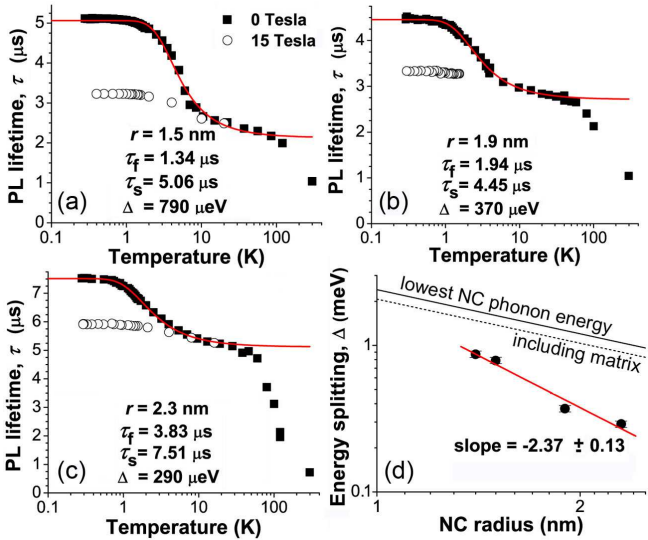}
\caption{(a-c) $\tau$ vs. $T$ for three PbSe NC sizes at $B$=0~T
(black points).  Lines are fits to a two-level model (see text).
Open symbols show $\tau (T)$ at $B$=15~T. High fields reduce $\tau$
to its value at $\sim$10~K. (d) The energy splitting $\Delta$
between these two lowest exciton levels versus NC radius.}
\label{fig2}
\end{figure}

Figs. 2(a,b) show $\tau(T)$ data for $r$=1.5 and 1.9~nm NCs, along
with similarly excellent fits of the low-$T$ data to the model. The
extracted splitting $\Delta$ decreases with NC size approximately as
$1/r^2$ [Fig. 2(d)], suggesting a quantum confinement origin. For
all NCs, $\Delta$ is much less than the smallest phonon energy
(which falls as $1/r$), even when including realistic acoustic
coupling to the surrounding organic matrix. Both $\tau_s$ and
$\tau_f$ tend to increase with NC size, although the correlation is
not strong.

We use high magnetic fields $B$ to further clarify this new exciton
fine structure. Figs. 2(a-c) also show $\tau(T)$ measured at 15~T.
For all NCs, the low-temperature lifetime is reduced to nearly its
10-50~K value, while for $T$$>$10~K, $\tau$ is unchanged by field.  The total PL intensity is unaffected by field.
These data suggest that the lowest-energy exciton gains some
oscillator strength due to field-induced mixing with the slightly
higher-lying state. That $B$ reduces $\tau$ to its 10-50 K value,
rather than its much smaller 300~K value, further confirms that
$\tau$ at $\sim$50~K reflects an intrinsic radiative recombination
time in PbSe NCs.

\begin{figure}[tbp]
\includegraphics[width=.48\textwidth]{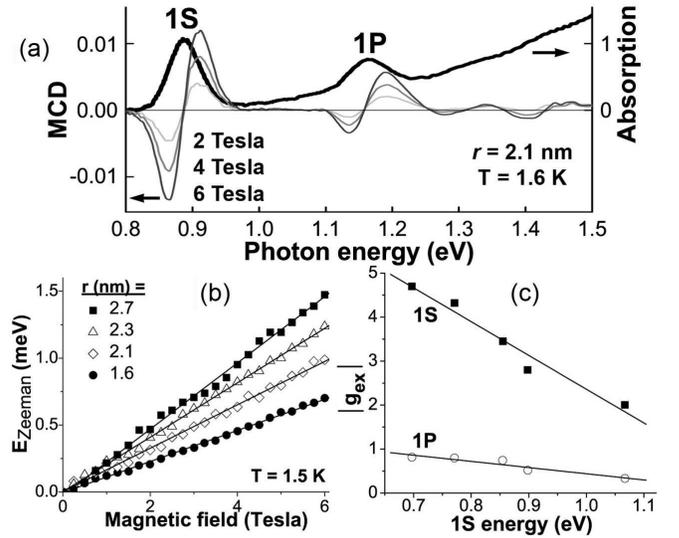}
\caption{(a) Optical absorption (right axis) and corresponding
magnetic circular dichroism (MCD; left axis) from \emph{r}=2.1~nm
PbSe NCs. (b) Zeeman splitting of the \emph{1S} absorption
feature, for different NCs. (c) The effective exciton
\emph{g}-factor $|g_{ex}|$ versus \emph{1S} band-gap energy.}
\label{fig3}
\end{figure}

An understanding of magnetic field effects requires some knowledge
of the (heretofore-unknown) magnetic energies of excitons in PbSe
NCs. To this end, magnetic-circular dichroism (MCD) studies were
performed (following Ref. \cite{Bussian}) in a 7~T split-coil magnet
using films of NCs diluted in poly(methyl methacrylate) to minimize
optical scatter. MCD, being a polarization-resolved absorption
measurement, measures primarily those excitons within the fine
structure that have large oscillator strength and which are
responsible for absorption (these excitons are typically higher in
energy than the emitting excitons from which PL originates, leading
to the Stokes' shift observed in PbSe and other NCs). We
measure the Zeeman splitting $E_Z$=$g_{ex}\mu_B B$ between right-
and left-circularly polarized optically-active excitons that are
split by a magnetic field. Fig. 3(a) shows the clear \emph{1S}
absorption of $r$=2.1~nm NCs, and the derivative-lineshape MCD
spectra, establishing that the \emph{1S} absorption is dominated by
an exciton having Zeeman-type splitting.  Fig. 3(b) shows
$E_Z(B)$ for different NC sizes, from which the \emph{g}-factors of
these absorbing excitons $|g_{ex}|$ are determined. $|g_{ex}|$
increases from 2 to 5 from smallest to largest NCs, which likely
results from increased spin-orbit coupling with decreasing band gap
\cite{Roth}.  Although MCD does not explicitly reveal the Zeeman
energy of the low-energy emitting states in the fine structure,
these studies do provide a clear indication of the relevant magnetic
energy scales in PbSe NCs.

The detailed dependence of $\tau$ on $B$ may now be considered. Fig.
4 shows $\tau(B)$ at 1.5~K for both large and small PbSe NCs. In an
applied field, $\tau$ at low temperatures is reduced by an amount
that depends on the degree of mixing between the lower- and
higher-energy exciton levels, which scales with the ratio of the
characteristic magnetic energy $E_Z$ to the level splitting
$\Delta$. For large PbSe NCs where MCD reveals large exciton
magnetic energies and where $\Delta$ is smallest, $\tau$ clearly
begins to saturate at high fields when $E_Z > \Delta$. The converse
holds for small NCs: $E_Z$ is small and $\Delta$ is big, and
$\tau(B)$ does not show saturation even by 15~T.  These data can be
roughly modeled by adapting the theory in Ref. \cite{Zeke}, which
described $B$-dependent mixing between bright and dark excitons in
CdSe NCs and its influence on $\tau$.

\begin{figure}[tbp]
\includegraphics[width=.48\textwidth]{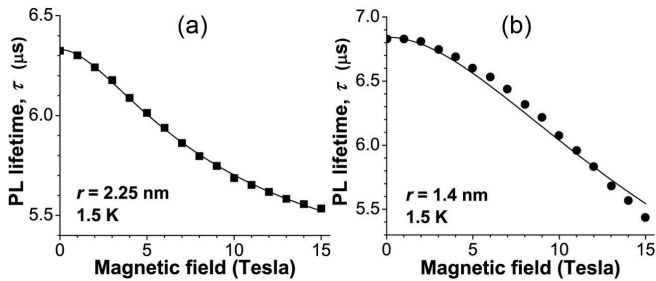}
\caption{Magnetic fields reduce $\tau$ at low $T$ (1.5~K), for both
(a) large and (b) small PbSe NCs. In large NCs, where Zeeman
energies are large but $\Delta$ is small (see text), $\tau$ begins
to saturate above 10~T. Lines show fits to the model of
\cite{Zeke}.} \label{fig4}
\end{figure}

Finally, magneto-PL studies show $>$10\% circularly-polarized
infrared emission at 2.5~K and 7~T [Fig. 5(a)], clearly implicating
the role of emitting excitons in the fine structure that are
Zeeman-split by magnetic fields and which couple selectively to
right- or left-circularly polarized ($\sigma^\pm$) light. As shown
in Figs. 5(b,c), the polarization $P$=$(I^+ - I^-)/(I^+ + I^-)$
increases with field (to 7~T) and decreases with temperature (to
40~K), in qualitative agreement with a thermal population of
excitons occupying or mixing with Zeeman-split levels. However,
since the precise symmetry, degeneracy and mixing character of the
emitting excitons are not established, $P$ cannot be used to
quantify the relevant \emph{g}-factors of these emitting states at
this time.

In summary, studies of PL and PL decay to 270~mK and 15~T provide
compelling experimental evidence for a new type of exciton fine
structure in rocksalt PbSe NCs that is qualitatively different from
that found in traditional wide-bandgap NCs such as CdSe. The data
indicate two weakly-emitting exciton levels with similar oscillator
strengths that are nearly degenerate in energy ($\Delta$$<$1 meV),
and which mix in magnetic field.  We also find that $\tau$ at 300~K
is likely dominated by non-radiative channels, such that the actual
radiative lifetime of excitons is much longer than previously
thought, of order 5 $\mu$s.  MCD studies establish the magnetic
(Zeeman) energy scale of absorbing states in the exciton fine
structure, and circularly polarized infrared magneto-PL is observed
for the first time. We anticipate that single-NC or resonant
(line-narrowing) studies will further quantify the precise ordering,
symmetry, and mixing of levels within the exciton fine structure of
PbSe and other lead-salt NCs.

\begin{figure}[tbp]
\includegraphics[width=.48\textwidth]{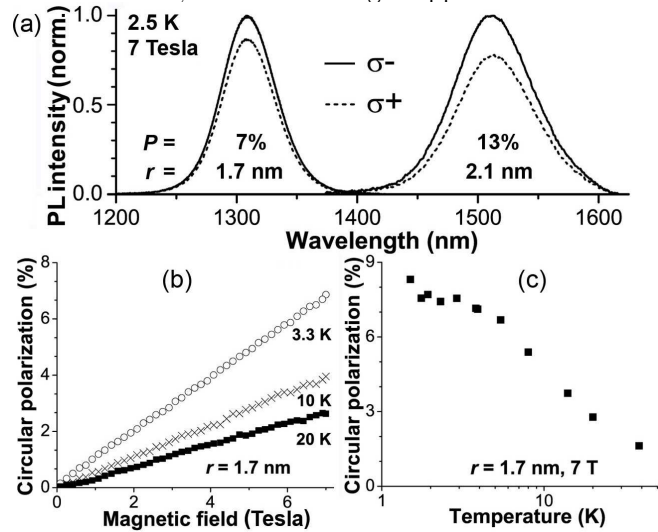}
\caption{(a) Circularly-polarized infrared PL emission from PbSe NCs
at $B$=7~T and $T$=2.5~K. (b) PL polarization versus $B$. (c) PL
polarization vs. $T$ at $B$=7~T.} \label{fig5}
\end{figure}
We thank A. Efros, D. Smith, and A. Migliori for helpful
discussions, and acknowledge support from the U.S. DOE Office of
Basic Energy Sciences (BES) Chemical Sciences, Biosciences and
Geosciences Division. V.I.K. is supported by the Center for Advanced
Solar Photophysics, a BES Energy Frontier Research Center.

\end{document}